# Thermodynamic signature of a magnetic-field-driven phase transition in the superconducting state of an underdoped cuprate

J. B. Kemper[1,*], O. Vafek[1], J. B. Betts[2], F. F. Balakirev[2], W. N. Hardy[3,4], Ruixing Liang[3,4], D. A. Bonn[3,4], and G. S. Boebinger[1]

1. *Dept. of Physics and National High Magnetic Field Laboratory, Florida State University, Tallahassee, FL 32310, USA*
2. *National High Magnetic Field Laboratory, Los Alamos National Laboratory, Los Alamos, NM 87545, USA*
3. *Department of Physics and Astronomy, University of British Columbia, Vancouver, British Columbia, V6T 1Z4, Canada*
4. *Canadian Institute for Advanced Research, 180 Dundas Street West, Suite 1400, Toronto, Ontario, M5G 1Z8, Canada*

**More than a quarter century after the discovery of the high temperature superconductor (HTS) $YBa_2Cu_3O_{6+\delta}$ (YBCO)** [1]**, studies continue to uncover complexity in its phase diagram. In addition to HTS and the pseudogap** [2, 3]**, there is growing evidence for multiple phases with boundaries which are functions of temperature (*T*), doping (p), and magnetic field** [4, 5, 6, 7, 8]**. Here we report the low temperature electronic specific heat ($C_{elec}$) of $YBa_2Cu_3O_{6.43}$ and $YBa_2Cu_3O_{6.47}$ (p=0.076 and 0.084) up to a magnetic field (*H*) of 34.5 T, a poorly understood region of the underdoped *H-T-p* phase space. We observe two regimes in the low temperature limit: below a characteristic magnetic field $H'\approx$12-15 T, $C_{elec}/T$ obeys an expected $H^{1/2}$ behavior** [9, 10]**; however, near $H'$ there is a sharp inflection followed by a linear-in-*H* behavior. $H'$ rests deep within the superconducting phase and,**

* Please send correspondence to: jonathon.kemper@gmail.com

**thus, the linear-in-*H* behavior is observed in the zero resistance regime** [11]**. In the limit of zero temperature, $C_{elec}/T$ is proportional to the zero-energy electronic density of states. At one of our dopings, the inflection is sharp only at lowest temperatures, and we thus conclude that this inflection is evidence of a magnetic-field-driven quantum phase transition.**

In elemental metals, the total low temperature specific heat has the well-known form $C=\gamma T+\beta T^3$, a sum of a linear term due to the electrons ($C_{elec}$) and a cubic term from the phonons. For ideal, perfectly clean $d_{x^2-y^2}$ superconductors, this form must be modified to $C(H=0,T)=\alpha T^2+\beta T^3$ due to the linear electronic DOS that arises from the presence of symmetry enforced nodal lines along which the momentum dependent superconducting gap, $\Delta_{\mathbf{k}}$, vanishes linearly with slope $d\Delta_{\mathbf{k}}/dk= v_\Delta$ [12]. Experimental measurements of the low temperature specific heat of HTS cuprates in magnetic fields applied along the c-axis have found that $C_{elec}(H,T)=\gamma(H)T$, [13, 14] consistent with theoretically predicted scaling for d-wave quasiparticles experiencing an "orbital" Doppler shift from superconducting vortices: $\gamma(H)-\gamma(H=0)\sim H^{1/2}/v_\Delta$ [9].

We have measured the specific heat of samples of $YBa_2Cu_3O_{6.47}$ (YBCO6.47) with $T_c$=49 K and YBCO6.43 $T_c$=41 K corresponding to hole dopings p=0.084 and 0.076, respectively [15]. The total specific heat (C) for $H< H'$ strongly resembles previously reported results for YBCO6.56 [16]. That is, for YBCO6.47, between 1 K and 8 K, we find $C_{elec}(H< H')/T= \gamma(H=0)+A_c H^{1/2}$ with $H'$=12 T, $\gamma(H=0)$=2.1 mJ mol$^{-1}$ K$^{-2}$ (Fig. 1) and $A_c$=0.6 mJ mol$^{-1}$ K$^{-2}$ T$^{1/2}$ (Fig. 2). The phonon term, $\beta$=0.38 mJ mol$^{-1}$ K$^{-4}$, varies negligibly (<1%) over the full range of $H$ (Fig. 1). Our second doping, YBCO6.43, also approximately follows $C_{elec}(H < H')/T = \gamma(H=0)+A_c H^{1/2}$ below 3 K, with $H'$=12 T, $\gamma(H=0)$=2.5 mJ mol$^{-1}$ K$^{-2}$ (Fig. 1) and $A_c$=0.6 mJ mol$^{-1}$ K$^{-2}$ T$^{1/2}$(Fig. 2). The data from YBCO6.43 in Figs. 1(d) and 2(b) are more temperature dependent than the data for YBCO6.47 in Figs. 1(b) and 2(a), yet they still follow the more general scaling law for d-wave superconductivity (SC) [10] below $H'\approx$15 T (Fig. 3). For both YBCO6.43 and YBCO6.47, the values of $A_c$ are within experimental error of the value measured for YBCO6.56 of $A_c$=0.57 mJ mol$^{-1}$ K$^{-2}$ T$^{1/2}$ [16]). At an additional intermediate doping of p=0.097, YBCO6.51, we have measured similar samples up to 15 T and also find $\gamma$(H=0)=2.3 mJ mol$^{-1}$ K$^{-2}$ and $A_c$=0.64 mJ mol$^{-1}$ K$^{-2}$ T$^{1/2}$ . We infer from these $A_c$ values that $v_\Delta \approx$0.15-0.17 eV Å [17, 18] and is insensitive to doping, despite the fact that the onset of finite resistance occurs at a magnetic field ($H_R$) that is a factor of ~1.5 *higher* in YBCO6.43 and YBCO6.47 than YBCO6.56 at low temperatures [11] The value of $v_\Delta$ in YBCO for p=0.08-0.10 is in strikingly similar to two other materials at the same dopings: in $Bi_2Sr_2CaCu_2O_{8+\delta}$ (BSCCO), as shown by tunneling (STM) [19] and photoemission (ARPES) ($v_\Delta \approx$0.11-0.12 eV Å) [20] as well as values of $A_c$ from specific heat of $La_{2-x}Sr_xCuO_4$ [21], from which we determine $v_\Delta \approx$0.16 eV Å [17, 18]. This constant magnitude of $v_\Delta$ for underdoped YBCO with $T_c$ ranging from 41 K to 59 K, which is similar to two other cuprates with $T_c$ varying from 20K (underdoped LSCO) [17, 18] to 92 K (near-optimum-doped BSCCO) [20], strongly implies that the pairing scale (inferred from $v_\Delta$) does not determine $T_c$ for underdoped cuprates. $v_\Delta$ is relatively constant over the range of dopings recently studied with specific heat (p=0.075-0.1) and yet roughly a factor of two larger than $v_\Delta$ measured at optimal doping, despite $T_c$ in the latter being over 90 K. That is, the energy gap is larger even though $T_c$ is lower in the underdoped samples studied here.

We now note that the predicted $H$=0 superconducting term C~$\alpha T^2$ is not clearly evident, however we can conclude $\alpha$< 0.2 mJ mol$^{-1}$ K$^{-3}$ (Methods). We note further that other specific heat [13, 14] and ARPES [22] measurements are consistent with this upper bound, given that $\alpha=18\,\zeta(3){k_B}^3 n_l ab \frac{1}{\pi\hbar^2 v_F v_\Delta}$ [12], where $v_F$ is the Fermi velocity, $\zeta(x)$ is the Riemann zeta function, $ab$ is the area of the a-b plane per unit cell, and $n_l$=2 is the number of $CuO_2$ layers *per mole* (

The most striking feature uncovered by the present study is the obvious deviation from $\Delta C_{elec}=C_{elec}(H,T)-C_{elec}(H=0,T)\sim H^{1/2}$ (see Methods) characterized by a low temperature inflection point and subsequent $\Delta C_{elec}$ ~$H$ behavior as seen in Figure 2. This approximately linear-in-$H$ behavior, unprecedented for HTS cuprates, leads to $\Delta C_{elec}$(H=34.5 T)/$T \approx$ 5-6 mJ mol$^{-1}$ K$^{-2}$ (Fig. 1b), roughly twice the value observed in YBCO6.56 at any $H \leq$ 45 T [16]. Two key facts are evident: (1) a larger specific heat exists at the lower dopings, YBCO6.43 and YBCO6.47, than at the higher doping, YBCO6.56. This is perhaps counterintuitive because SC is more robust at the lower dopings, thus one might expect a larger superconducting gap which would ordinarily lead to a smaller specific heat. That is, at $T\approx$ 1 K, $H_R\approx$45 T for YBCO6.43 and YBCO6.47, while $H_R$<35 T for YBCO6.56 [11, 16] (2) In YBCO6.56, over any portion of the magnetic field range measured, all evidence points to the absence of linear-in-$H$ enhancement of $C_{elec}$ above the $H^{1/2}$ fit extrapolated from low-field data [16]. Moreover, $\Delta C_{elec}/T$ is essentially temperature independent in YBCO6.47, even for data crossing into the resistive regime at $H_R(T)$, indicated by the arrows in Fig. 1b. [11] The magnetic field dependence of $\Delta C_{elec}/T$ is also not affected by $H_R(T)$. These facts, taken together make it unlikely that $H'$ and $H_R(T)$ mark the mean-field suppression of the superconducting gap by magnetic field. In addition, the abruptness of the change at $H'$ is at odds with the smoothness expected in the magnetic field suppression of the order parameter of a d-wave superconductor [23].

In Fig. 3, we re-plot the YBCO6.43 data to test for a more general d-wave scaling (SL scaling) predicted by Simon and Lee [10]. All data below $H'$=15 T scale within the scatter between 1 and 7 K. The breakdown in scaling above 15 T is most clearly visible in the dramatic upturn in the 2.5 K trace (Fig 3c) which represents a sharp deviation from scaling. Note that all data for $H\geq$18 T are above $H'$, and do not scale. A broad maximum centered near 1 K/$T^{1/2}$ exists for all fields. The quotient $T/H^{1/2}$ is proportional to the ratio of the orbital magnetic length over the thermal length, suggesting the maximum may arise from superconducting vortices: in fact, a similar anomaly in the low temperature $C_{elec}$ of d-wave superconductors has been predicted as a consequence of magnetic sub-bands resulting from the periodicity of the vortex lattice [24].

We re-plot the ~2 K data of Fig. 2 (a) in Fig. 4(a) to demonstrate the breakdown of orbital scaling at $H'$, and to clearly illustrate the sharpness of the transition between the two regimes. The plot depicts a scenario in which the orbital effect, $\Delta C_{elec}(H)/T \propto H^{1/2}/v_\Delta$, is responsible for the observed $\Delta C_{elec}(H)/T$ over the entire magnetic field range. Such a scenario would necessitate a field-dependent $v_\Delta^{eff}(H) \propto TH^{1/2}/\Delta C_{elec}(H)$ (where $v_\Delta^{eff}$ is an effective field-dependent parameters such that for $H \rightarrow 0$, $v_\Delta^{eff} = v_\Delta$) that is more or less a piecewise function: $v_\Delta^{eff}$ is largely field-independent below $H'$, followed by a sudden drop at $H'$ that asymptotes to $v_\Delta^{eff} \propto 1/H^{1/2}$ (Fig. 4a). Since $v_\Delta^{eff}$ is proportional to the magnitude of the superconducting order parameter, this further disfavors a mean-field scenario involving suppression of the gap by $H$ [23] .

Instead, it is natural to think of Zeeman splitting giving rise to linear-in-$H$ specific heat, which can result from Zeeman splitting of d-wave quasiparticles at sufficiently high fields [25]. In such case, the high field slope $\Delta C_{elec}(H)/T\,H \sim (v_F v_\Delta)^{-1}$, and, assuming a g-factor of 2, our value for this slope, $\approx$0.16 mJ mol$^{-1}$ K$^{-2}$ T$^{-1}$, determines $\alpha \approx$ 0.37 mJ mol$^{-1}$ K$^{-3}$ [12, 25]. This violates our established zero-field bound on $\alpha$ by nearly a factor of 2, and requires a dramatic drop in the product $v_F^{eff} \times v_\Delta^{eff}$ over a small field range around $H'$ for this Zeeman scenario to be internally consistent.

Fig. 4b presents this second scenario, with fixed $v_\Delta$ (determined from $A_c$, the prefactor of the square root H) and a field-dependent $v_F^{eff}$ calculated to fit the low temperature data. The magnitude of $v_F^{eff}$ decreases by at least a factor of 3 from its low field value, suggesting the sudden onset of a mass enhancement at $H'$. The phase diagram of YBCO (Fig. 5) supports this picture and provides a scenario for understanding the observed phase transition: In the resistive state using pulsed magnetic fields above 50 T, quantum oscillations (QO) [26] and resistivity measurements [27] have been interpreted in terms of a metal to insulator transition (MIT) near YBCO6.47with a divergent cyclotron mass, m*[26]. In light of Fig. 4b, we propose that this enhanced mass is a field sensitive phenomenon, which increases rapidly at $H'$ and drives the system into a Zeeman-dominated regime. We note that $v_F^{eff}/v_\Delta \approx 2/3$ above 20 T, a violation of the requirement that $v_F >> v_\Delta$ in the orbital scaling regime, providing a further quantitative indication that the high-field regime is physically distinct from the low-field.

A proposed quantum critical point at p=0.08 [27, 28] linked to the high-field mass enhancement [26] has been associated in YBCO with the boundary of a spin density wave phase (SDW) seen via neutron diffraction and muon spin rotation (**μSR)** [5, 6] (Fig 5), whose order is enhanced by magnetic field [6]. Our high field state may be linked to this SDW and its associated critical point. Indeed, a field-driven transition to a SDW state -coexisting with SC- has been reported in the cuprates, although this observation was in a different material ($La_{2-x}Sr_xCuO_4$) at a much higher doping (p=0.14) [29]. Unfortunately, the most comprehensive survey of $La_{2-x}Sr_xCuO_4$ specific heat [21] does not include a sample at a doping that would traverse this reported phase boundary. A neutron scattering experiment in YBCO at these dopings and magnetic fields might provide further evidence of SDW order underlying the magnetic-field-driven phase transition that we report here.

**Methods**

The YBCO samples included multiple crystals, all with nearly identical doping and total mass 6.0 mg. Initially, these detwinned crystals had δ=0.51 (YBCO6.51, $T_c$=57 K, p=0.097) [15], and we measured C up to 15 T. The samples were then re-annealed and detwinned at δ=0.47 (YBCO6.47, $T_c$=49 K, p=0.84) [15], and measured again; the process was then repeated to reach δ=0.43 (YBCO6.43, $T_c$=41 K, p=0.76) [15] before the final measurements. The $T_c$'s were established via MPMS as the midpoint of the magnetization transition with applied field ($H$<2 Oe) parallel to c-axis. The $H$=0 specific heats of each of three YBCO6.47 samples were measured individually and were identical. Measurements of YBCO6.47 in finite field were carried out on both a two sample and a three sample aggregation, 3.9 mg and 5.96 mg respectively; these two sets of measurements agree closely at all temperatures. All specific heat data was acquired by a high-speed modified relaxation time method.

Progression through the *H-T* plane during data acquisition employed two modes: "Field fixed" mode is defined as moving through the temperature range 1- 8 K by changing the temperature set-point while the field remained fixed between data collections; "Field step" involved the choice of single temperature set-points and ramping of the field between data collections. A single "data collection" is defined as one continuous accumulation of data at a single field/temperature set-point. Before any and all changes magnetic field, the samples were heated into the resistive state and held there until the field was fixed to a new set point. The samples were subsequently "field cooled" to a temperature set-point before beginning data collection.

The extraction of $\Delta C_{elec}(H,T)$ from the total C involves a precise process of subtractions of well-characterized extraneous contributions, including the heat capacity of the apparatus, or addenda. The relation between these contributions and the total is a simple sum, $C(H,T)=C_{elec}(H,T)+\beta T^3+C_{Schottky}+C_{hyp}+O(T^5)$ with each term visible in Fig. 1(a) and Fig. 1(c). $C_{Schottky}$ and $C_{hyp}$ are defined in [13] and the supplementary information from ref. [16]. $C_{Schottky}+C_{hyp}$ is negligible for $C(H=0, T>3$ K) and $C(H=34.5$ T, $T>3$ K) allowing us to establish a β that varies by less than 1% between $H=0$ and 34.5 T, as shown in Fig 1 (a) and (c). Taking β as fixed, we determine $C_{Schottky}$ above 3 T to be due to a Zeeman split localized electron spin with g-factor=2. Below 3 T, $C_{Schottky}$ shows a well-known $H=0$ broadening [13], which makes subtraction difficult, possibly due to interactions or a small internal field which our analysis implies must be perpendicular to $H$. The data above 2 T is well-behaved, and from it we estimate a reasonable density of contributing spins for YBCO6.47, about 1 per 300 unit cells, roughly 50% larger than the value reported for YBCO6.56 [16]. Further, the density of spins is about 15% larger in YBCO6.43 than YBCO6.47, while that for YBCO6.51 is slightly larger than that for YBCO6.56, indicating a possible link between $C_{Schottky}$ and oxygen vacancies. $C_{hyp}$ is treated with a parabolic interpolating function $F(H)$ where $C_{hyp}=F(H)/T^2$. The function can be approximated by $F(H)=PH^2$ with $P=0.0094$ +/- 0.0003 mJ mol$^{-1}$ T$^{-2}$, similar to Riggs *et al.* [16] which reported $P=0.011$ +/- 0.002 mJ mol$^{-1}$ T$^{-2}$.

The technique for acquiring data is a modified relaxation time method where a single heat capacity datum is determined by dividing the maximum heater power during a square heating pulse by the difference of warming and cooling rates at a single temperature. The warming takes place over a small temperature interval, typically between 30-100 mK, and is driven with a measured, constant DC power Q. The rate of temperature increase is determined with a field-calibrated resistive thermometer attached to the sample, and the cooling rate is measured subsequently after Q is set to 0. The actual datum is an effective average of $Q/(dT/dt|_{warming} - dT/dt|_{cooling})$ where t is the time. The method is checked for accuracy by varying Q, the time over which Q>0, and varying the current applied in the thermometer measurement. The profile of the warming and cooling curves match that of a system where the heater and thermometer are in thermal equilibrium with the sample and well isolated from the ambient environment.

Typical time intervals for a single warming and cooling pair span 0.25 to 5 seconds for a 4 mg sample. This time depends on the total C, including addenda, as well as the thermal conductance into the environment. A data point with error bars in any figure is generated from a single “data collection”, defined as a series of C data points taken one immediately after the other at a single ($H$,$T$) setpoint. Typical data collections include 5-20 points, with the error bars spanning +/− one standard deviation. This definition of error bar does not apply to fit parameters, which are given as +/− 2 standard errors as determined by a fitting routine, unless otherwise noted. Except where noted, all measurements were performed with the magnetic field vector **H** parallel to the crystalline c-axis.

**Acknowledgements** The authors thank Steven Kivelson, Ryan Baumbach, Aharon Kapitulnik, Mike Norman, Brad Ramshaw, Arcadi Shekhter, Jeff Sonier, and Scott Riggs for discussion and commentary on the manuscript, as well as Albert Migliori for discussions on the experimental technique. J.B.K. thanks Camilla Moir for assistance during experiments. A portion of this work was performed at the National High Magnetic Field Laboratory, which is supported by National Science Foundation Cooperative Agreement No. DMR-1157490, the State of Florida, and the U.S. Department of Energy. Work at the University of British Columbia was supported by the Natural Science and Engineering Research Council of Canada and the Canadian Institute for Advanced Research.

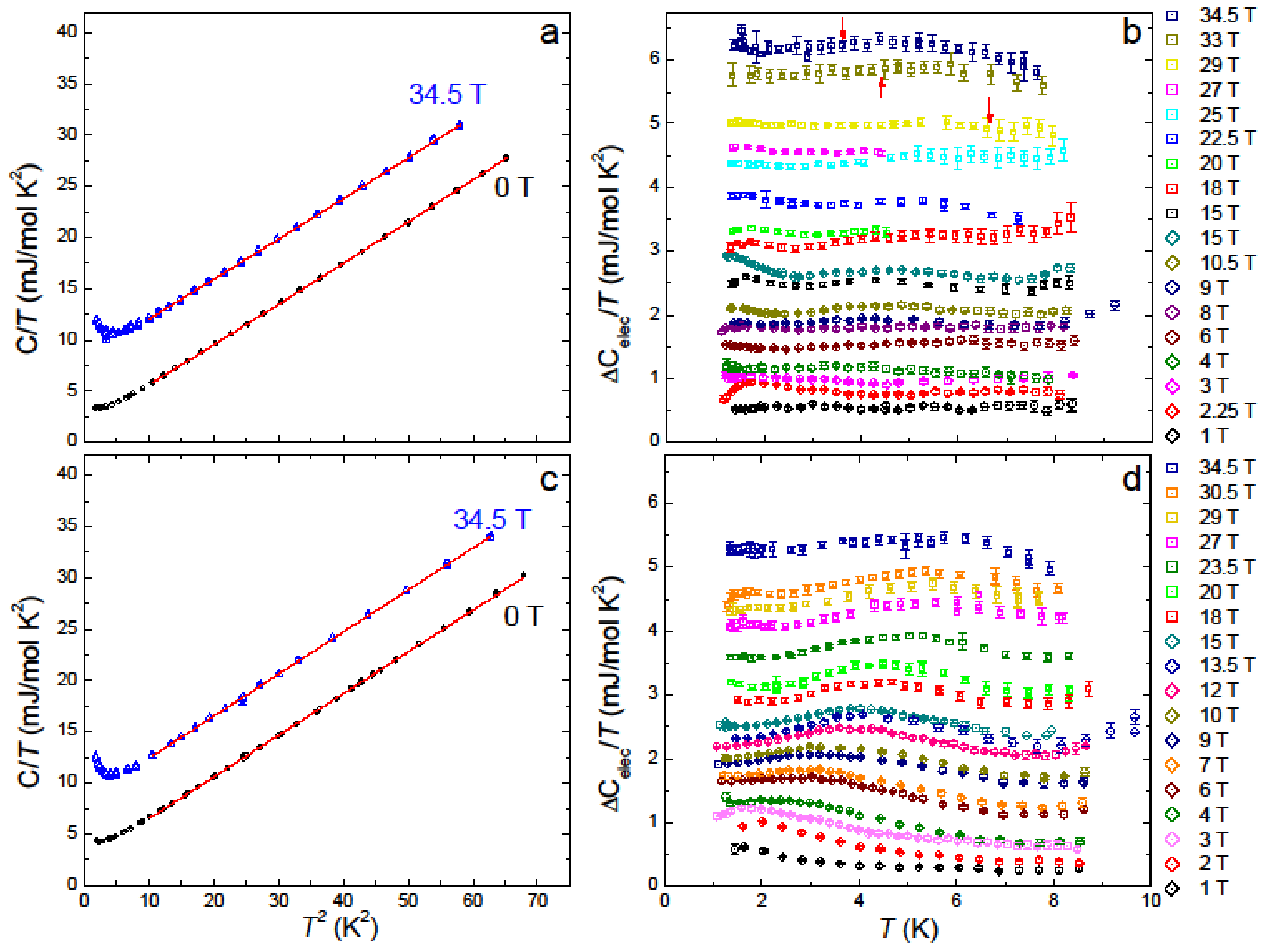

**Figure 1 Temperature dependence of the specific heat.** (a) YBCO6.47 Total C/$T$ versus $T^2$ at $H$=0 T (black) and 34.5 T (blue) with linear fits (red) to determine $\gamma(H=0)=2.2$ mJ·mol$^{-1}$·K$^{-2}$, $\beta(H=0)=0.381$ mJ·mol$^{-1}$·K$^{-4}$, and $\beta(H=34.5\text{ T})=0.382$ mJ·mol$^{-1}$·K$^{-4}$ (b) $\Delta C_{elec}/T$ versus $T$ at various fields. (c), (d) Same plots for YBCO6.43. Fits in (c) give $\gamma(H=0)=2.5$ mJ·mol$^{-1}$·K$^{-2}$, $\beta(H=0)=0.407$ mJ·mol$^{-1}$·K$^{-4}$, and $\beta(H=34.5\text{ T})=0.411$ mJ·mol$^{-1}$·K$^{-4}$. Error bars in plots are +/- one standard deviation for a single data collection (Methods). The arrows in panel (b) indicate the approximate position of the resistive transition for $H\geq29$ T as reported for YBCO6.47 [11].

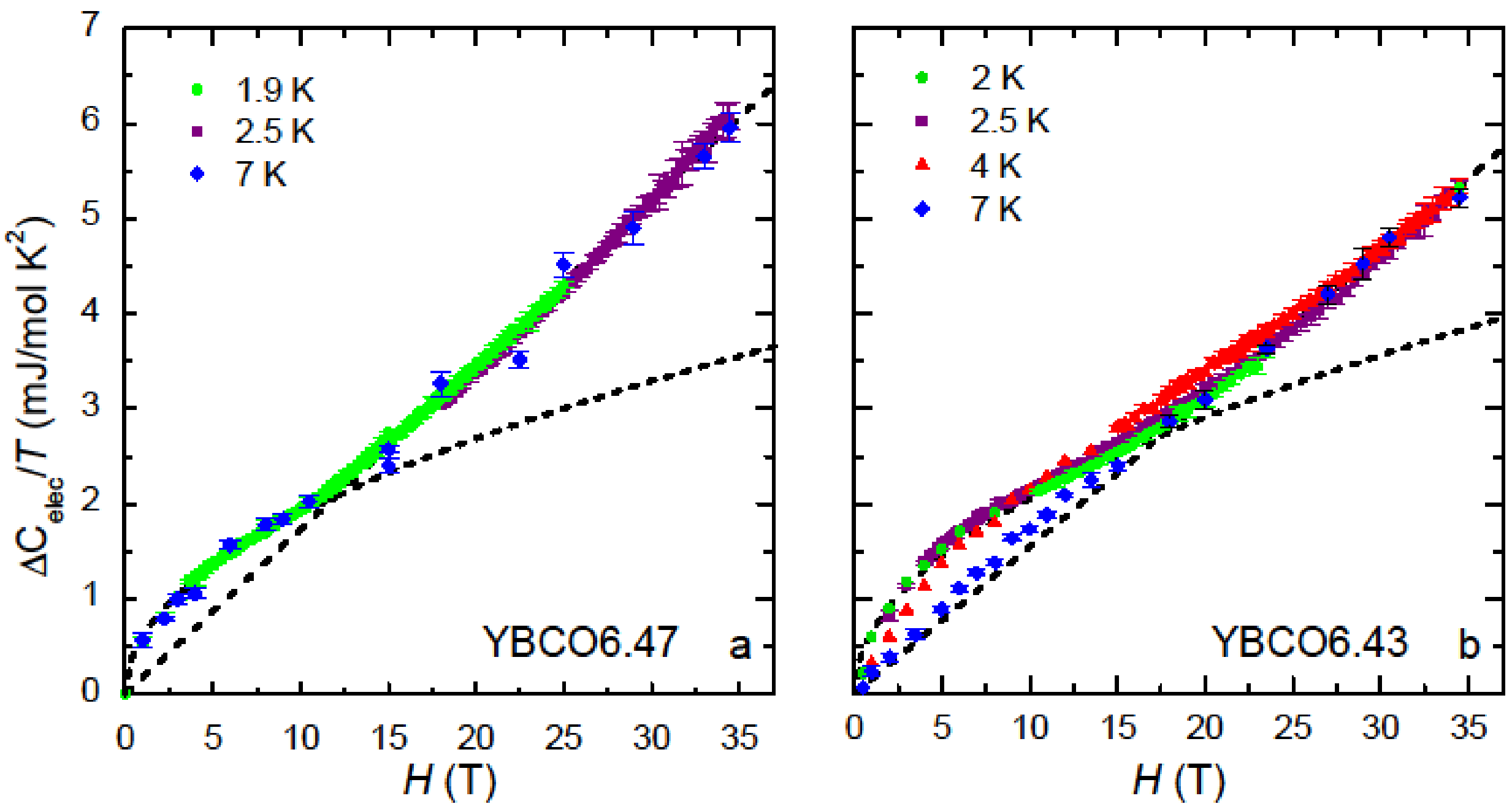

**Figure 2 Field dependence of electronic specific heat.** (a) YBCO6.47 $\Delta C_{elec}/T$ versus $H$. The dashed black curve and line are guides to the eye, given by $\Delta C_{elec}/T$=0.6 mJ mol$^{-1}$ K$^{-2}$ T$^{-1/2}$×$H^{1/2}$ and $\Delta C_{elec}/T$= 0.17 mJ·mol$^{-1}$·K$^{-2}$·T$^{-1}$×$H$, respectively, where $H$ is given in teslas. (b) YBCO6.43 $\Delta C_{elec}/T$ versus $H$. Again, the dashed curve and line are guides to the eye: $\Delta C_{elec}/T$=0.6 mJ·mol$^{-1}$·K$^{-2}$·T$^{-1/2}$×$H^{1/2}$ and $\Delta C_{elec}/T$= 0.155 mJ·mol$^{-1}$·K$^{-2}$·T$^{-1}$×$H$, respectively. Error bars in plots are +/- one standard deviation for a single data collection (Methods).

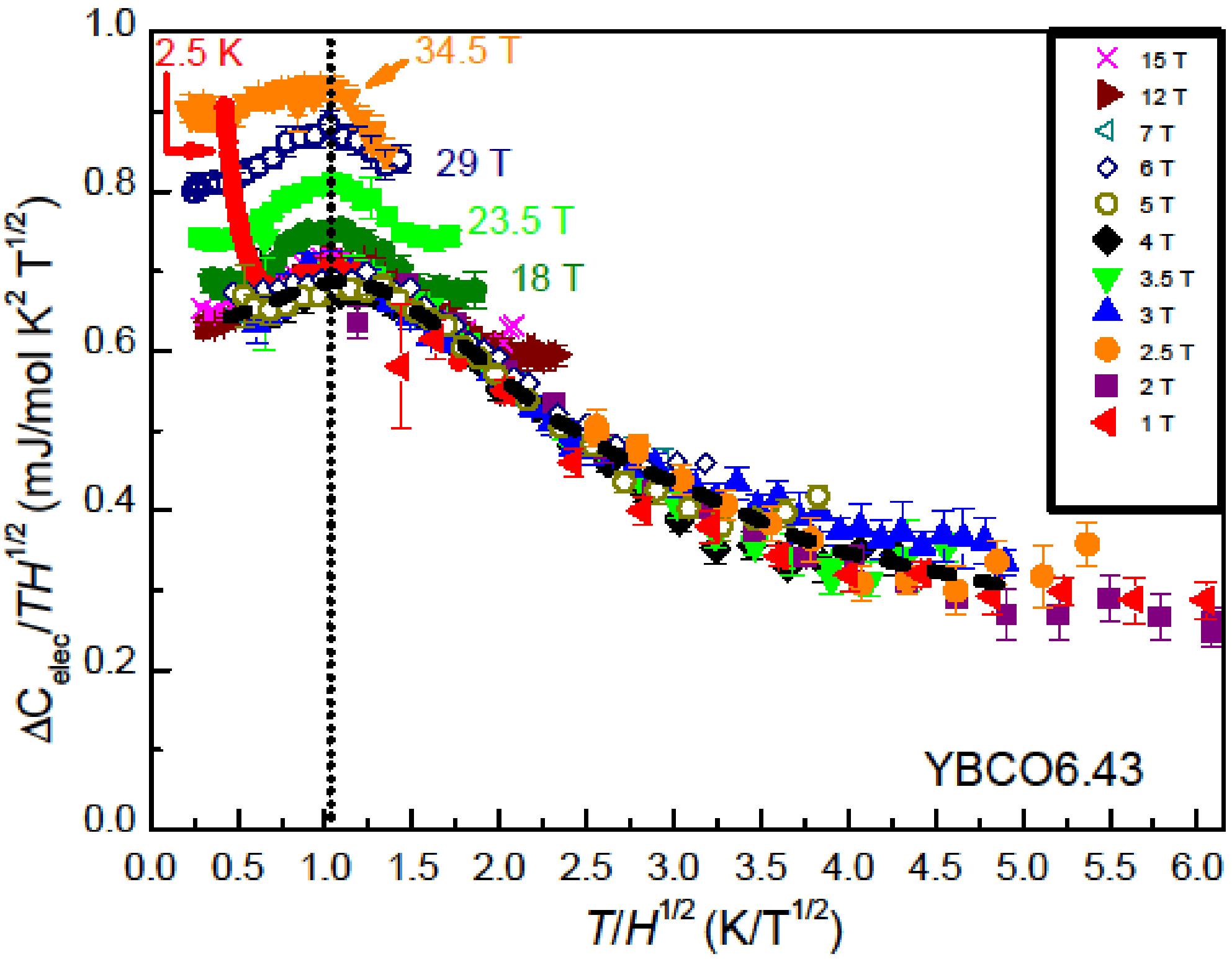

**Figure 3 Scaling plot.** Data from Fig.1(d) plus the 2.5 K data from Fig. 2(b) (red circles) plotted to test for SL scaling for d-wave SC [10]. The data collapse within error for T<=7 K, provided H<=15 T. The dashed black curve traces an approximate scaling function. The legend lists fixed-field curves that follow scaling below ≈7 K. The dotted vertical line indicates an anomaly consisting of a maximum at $T/H^{1/2}\approx 1$ K·T$^{-1/2}$ at all fields. The red arrow indicates the upturn in the 2.5 K data. Error bars in plots are +/- one standard deviation for a data collection.

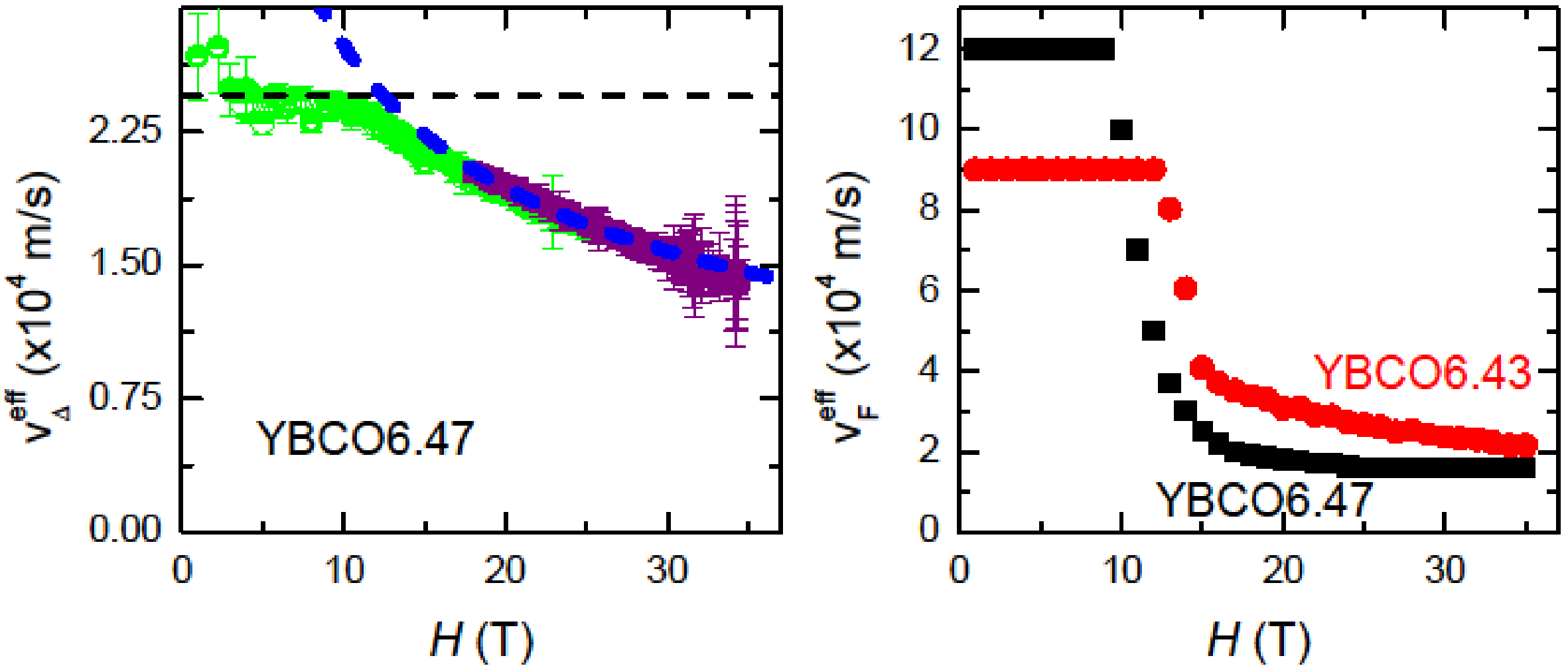

**Figure 4 Calculated effective electronic velocity parameters. (a)** Re-plotting of the data in Fig. 2 (a) as $v_\Delta^{eff}$ versus *H*, ($v_\Delta \propto TH^{1/2}/\Delta C_{elec}$ for ideal d-wave superconductors with constant $v_F^{eff}$). Color scheme is the same as Fig. 2(a), and dashed lines are again defined by $\Delta C_{elec}/T$=0.17 mJ·mol$^{-1}$·K$^{-2}$·T$^{-1}$×*H* and $\Delta C_{elec}/T$=0.6 mJ·mol$^{-1}$·K$^{-2}$·T$^{-1/2}$×$H^{1/2}$. Error bars represent +/- one standard deviation for one data collection, recalculated for the quantity$TH^{1/2}/\Delta C_{elec}$ (b) Alternative scenario assuming fixed $v_{\Delta,}$ showing calculated $v_F^{eff}$**[18]** with both orbital and Zeeman effects included. The value of $v_F^{eff}$ below *H′* is determined from thermal conductivity**[30]**. The drop in $v_F^{eff}$ implies an enhanced quasiparticle mass above *H′*.

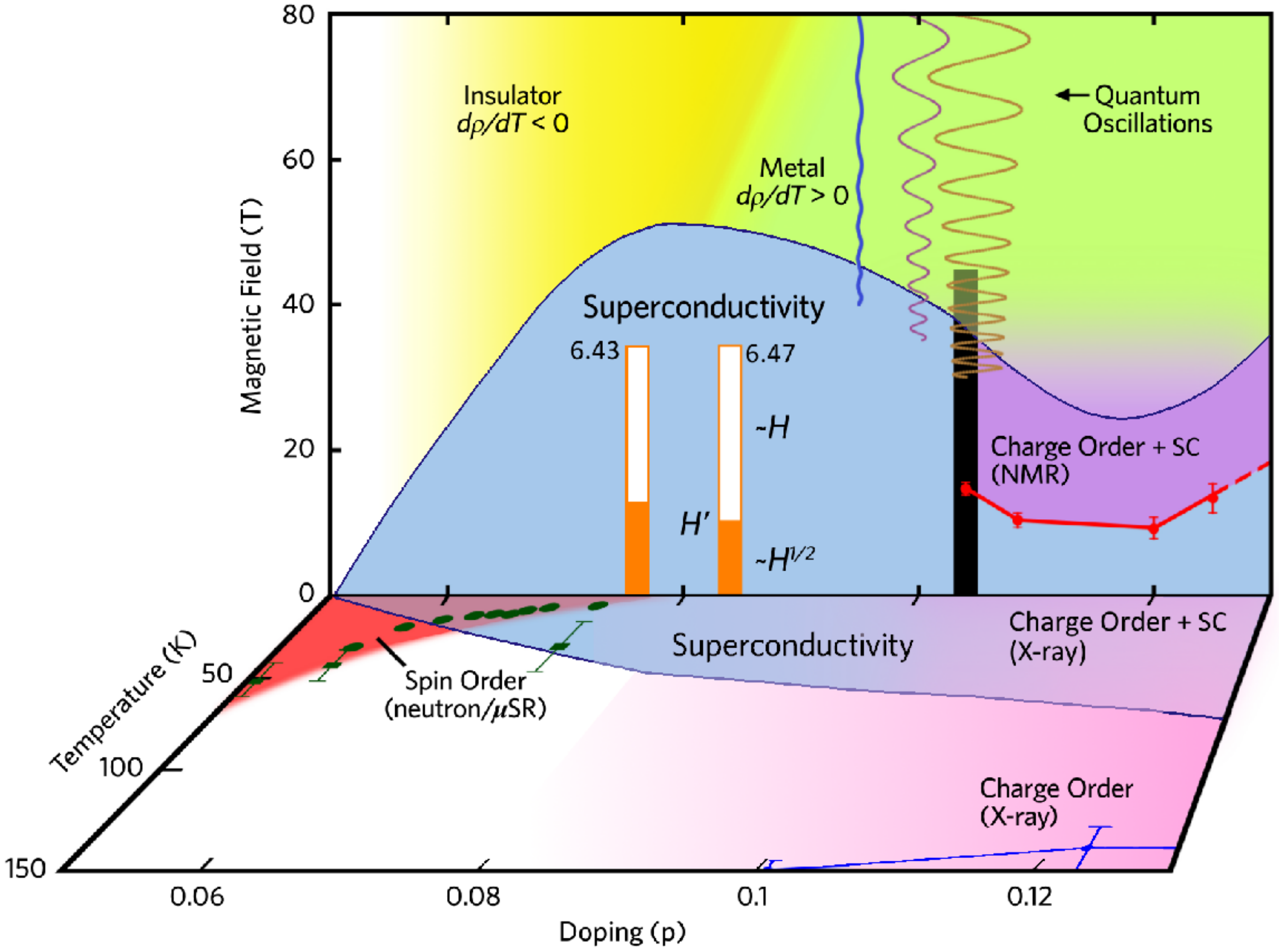

**Figure 5 Phase diagram.** Panels represent the *H*-p plane for $T \rightarrow 0$ and *T*-p plane for $H \rightarrow 0$. Orange/white rectangles represent the range of the present measurements, and the black rectangle represents ref. **[16]**. Orange and white represent the range of d-wave scaling (i.e. $C_{elec}/T \sim H^{1/2}$) and linear-in-*H* behavior, respectively. *H′* marks the phase transition. SC phase boundaries (blue curves) **[15][11]**, MIT (green/yellow boundary) from resistivity**[27]** and QO**[26]**, and CDW from NMR**[7]** (purple, red circle/line) and x-rays**[8, 30]** (pink, blue circle/line), are shown (CDW may extend below p=0.08**[28]**). SDW (red) at *H*=0 neutrons**[6]** (green squares) and μSR**[4, 5]** (green circles) vanishes near p=0.08.